\documentclass[12pt]{iopart}
\usepackage{verbatim}
\usepackage{iopams}

\usepackage{color}
\usepackage{algorithm}
\usepackage{algorithmic}
\usepackage{graphicx}

\def\FF{\mathbb{F}}
\def\bZ{\mathbb{Z}}
\def\sX{\mathsf{X}}

\def\eqref#1{(\ref{#1})}

\def\Label#1{\label{#1}\ [\ \text{#1}\ ]\ }
\def\Label{\label}

\begin{document}

\title[Secure Quantum Network Coding on Butterfly Network]{Secure Quantum Network Coding on Butterfly Network}

\author{Masaki Owari}
\address{Department of Computer Science, Faculty of Informatics, Shizuoka University}
\ead{masakiowari@inf.shizuoka.ac.jp}
\author{Go Kato}
\address{NTT Communication Science Laboratories, NTT Corporation}
\ead{kato.go@lab.ntt.co.jp}

\author{Masahito Hayashi}
\address{Graduate School of Mathematics, Nagoya University, \\
Centre for Quantum Technologies, National University of Singapore}
\ead{masahito@math.nagoya-u.ac.jp}

\vspace{10pt}
%\begin{indented}
%\item[]February 2014
%\end{indented}

\begin{abstract}
Quantum network coding on the butterfly network 
has been studied as a typical example of quantum multiple cast network.
We propose secure quantum network coding on the butterfly network in the multiple unicast setting based on a secure classical network coding.
This protocol certainly transmits quantum states when there is no attack.
We also show the secrecy even when 
the eavesdropper wiretaps one of the channels in the butterfly network.
\end{abstract}

\noindent{\it Keywords}: secrecy, quantum state, network coding, butterfly network, multiple unicast

\submitto{Quantum Science and Technology}
%\maketitle

\section{Introduction}
Construction of quantum network is one of next targets of quantum information processing.
For this purpose, several researchers \cite{Hayashi2007,PhysRevA.76.040301,Kobayashi2009,Leung2010,Kobayashi2010,Kobayashi2011}
have studied network coding for quantum network, which realizes efficient transmission of quantum state via quantum network.
Network coding has several formulations.
Most simple formulation is the unicast setting, in which,
we discuss the one-to-one communication via the network.
This formulation is discussed in many studies on the classical network coding.
Since conventional network has many users, we need to treat networks that has several users.
As such a formulation, we often focus on the multicast setting,
in which, one sender sends the information to plural receivers.
However, in the quantum setting, it is impossible due to the no-cloning theorem.
Hence, we discuss the multiple unicast setting, which has plural pairs of a sender and receiver.
As one of simplest examples of the multiple unicast setting, we often focus on the butterfly network.
Therefore, it is natural to consider the butterfly network in the framework of quantum network coding.

For example, the paper \cite{Hayashi2007} started a study of quantum network coding with the butterfly network.
The paper \cite{PhysRevA.76.040301} clarified the importance of the prior entanglement in quantum network coding in the case of the butterfly network.
The papers \cite{Kobayashi2009,Leung2010,Kobayashi2010,Kobayashi2011} generalized these results to more general settings.
As quantum information processing is related to secure protocols,
the security analysis is more needed for quantum network coding. 
Now, we consider the case when there is a possibility that an adversary might attack the quantum network.
In this case, we can guarantee the security in these existing codes if we verify the non-existence of the eavesdropper.
However, this method requires us to repeat the same quantum state transmission several times.
It is impossible to guarantee the security under the single transmission in the simple application of these existing methods 
because the verification requires several times of transmission.
Therefore, it is needed to propose a quantum network code to guarantee the security. 
For this purpose, we consider a natural extension of the methods of classical secure network coding.

Although the paper \cite{ACLY} started the study of network coding for the classical network, 
the paper \cite{Cai2002} initiated to address the security of network coding, and
pointed out that the network coding enhances the security.
Currently, many papers 
\cite{Cai2002,Cai06,Bhattad2005,Liu2007,Rouayheb2007,Harada2008,SK,Cai2011,Cai2011a,Matsumoto2011,Matsumoto2011a,KMU} 
have already studied the secrecy for network coding.  
These papers offers the security even when Eve eavesdrops a fixed number of links in the given network.
That is, the security is guaranteed whatever links are eavesdropped when the number of eavesdropped links is less than the given threshold.
Moreover, the multiple unicast setting has not been well examined even in the classical case, i.e., 
only a few papers such as Agarwal et al. \cite{Agarwal} discuss this setting with classical case.

The purpose of this paper is to propose a protocol to guarantee the security 
for the transmitted quantum state in the single-shot setting
under the butterfly network even when any link is eavesdropped
whenever the number of eavesdropped links is one.
This type code is a natural extension of the above classical secure network coding,
and realizes the security without verification.
To realize the above requirements,
our code needs an additional shared randomness in the sink side, which is not needed in the original classical network code.
Since quantum channel is much more expensive than classical public channel,
we assume that any amount of classical public channel is freely used.
Under this assumption, transmission of quantum state is equivalent to sharing maximally entangled state via quantum teleportation \cite{PhysRevLett.70.1895}.
So,  we prove that the entangled state is shared by sending entanglement halves from sink nodes via our protocol under the above assumption.

The remaining part of this paper is organized as follows.
Section \ref{S2} introduces several notations for the butterfly network of the multiple unicast setting, and a secure classical network coding protocol on the butterfly network.
In Section \ref{S3}, we explain our secure quantum network coding protocol.
Section \ref{S4} shows  the secrecy of the transmitted quantum states even when any one of the quantum channels in the butterfly network.

\section{Preparation and classical protocol}\Label{S2}
We focus on the quantum butterfly network based on the finite filed $\FF_p:=\bZ/ p \bZ$ with prime $p\ge 3$ as Fig. \ref{F1}.
%The purpose of this quantum communication network
The task is transmitting quantum states from two source nodes $V_1$ and $V_2$
to  two sink nodes $V_5$ and $V_6$ via the quantum communication network
composed of intermediate nodes $V_3$ and $ V_4 $ and the edges $e(5), \ldots, e(11)$, which correspond to 
quantum channels.
In this network, the two source nodes $V_1$ and $V_2$ share a common random number,
and two sink nodes $V_5$ and $V_6$ share another common random number.

In the quantum setting,  
we need to separately describe input quantum state and encoding operation.
%we discuss whether entangled state is shared by sending entanglement halves from sink nodes.
So, we treat virtually vertices that have only quantum system and does not have any operation.
Since we make quantum operations in both source nodes $V_1$ and $V_2$
and both sink nodes $V_5$ and $V_6$, 
we additionally prepare input vertices $I_1$ and $I_2$ and output vertices $O_1$ and $O_2$ as other edges.
Hence, inputting halves of entangled states in both input vertices $I_1$ and $I_2$,
we can check our protocol generates entangled states
between other entanglement halves and output vertices $O_1$ and $O_2$.

Therefore, we address the following quantum network.
The edges of this network are composed of input vertices $I_1$ and $I_2$,
output vertices $O_1$ and $O_2$,
classical shared randomness sources $S_1$ and $S_2$,
and nodes $V_1, \ldots V_6$.
The edges of this network are composed of $e(1), \ldots, e(15)$.
Here, only edges $e(3),e(4),e(14),e(15)$ are classical channels, and other edges are quantum channels.
The numbers assigned to edges express the time ordering of the transmission on the corresponding channel.

\begin{figure}[htbp]
\begin{center}
\scalebox{1}{\includegraphics[scale=0.5]{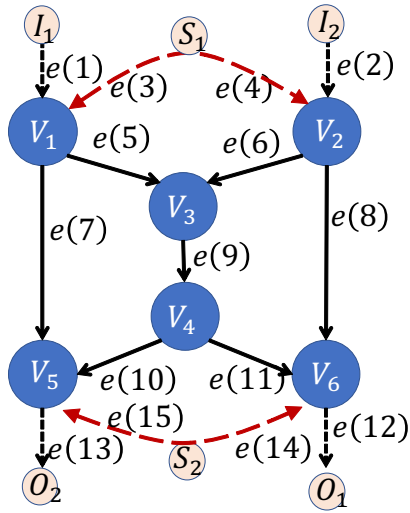}}
\end{center}
\caption{Butterfly network}
\Label{F1}
\end{figure}%

To give our quantum network coding, 
as a preparation, we proposed a specific type of a novel classical network coding of the multiple-unicast task in the butterfly network.
To express the information flow, we denote the information on the edge $e(i)$ by $Z_i$.
That is, the aim of our coding is to transmit the information from $V_1$ ($V_2$) to $V_6$ ($V_5$), respectively.
For this purpose, we employ the following coding on the respective edges of the network.
We assume the following information flow of the butterfly network.
\begin{eqnarray}
V_1 \qquad & Z_5:=2Z_1+Z_3, \quad Z_7:=Z_1+Z_3. \nonumber \\
V_2 \qquad &Z_6:=2Z_2+Z_4, \quad Z_8:=Z_2+Z_4. \nonumber  \\
V_3 \qquad &Z_9:=Z_5+Z_6. \nonumber  \\
V_4 \qquad  &Z_{10}:=Z_9, \quad Z_{11}:=Z_9. \nonumber \\
V_5 \qquad & Z_{13}:=\frac{1}{2}Z_{10}- Z_7.\nonumber \\
V_6 \qquad & Z_{12}:=\frac{1}{2}Z_{11}- Z_8.\label{O02}
\end{eqnarray}

Here, we denote the information to be sent from $I_i$ and the shared randomness generated in $S_i$
by $A_i $ and $B_i$, respectively.
So, we have $Z_1=A_1$, $Z_2=A_2$, $Z_3=Z_4=B_1$, $Z_{14}=Z_{15}=B_2$.
%we do not use the notations $Z_1,Z_2,Z_3,Z_4, Z_{14}, Z_{15}$.
In this classical setting, we do not use the second shared randomness $B_2$.
Therefore, we obtain the information flow with respect to the original information  $A_1 $, $A_2$, and $B_i$ in Fig. \ref{F2}.
So, we find that the sink nodes $V_6$ and $V_5$ correctly recover the information $A_1$ and $A_2$, respectively.
\begin{figure}[htbp]
\begin{center}
\scalebox{1}{\includegraphics[scale=0.5]{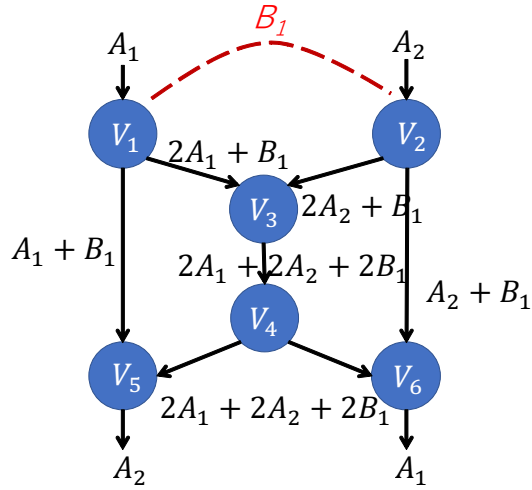}}
\end{center}
\caption{Information flow in butterfly network}
\Label{F2}
\end{figure}%
Thanks to the shared randomness, even though 
the eavesdropper, Eve, wiretaps one of edges $e(5), \ldots, e(11)$, 
she cannot obtain any information for $A_1$ and $A_2$.

Since the resultant values $Z_{1}, \ldots,  Z_{13}$ are determined by $A_1$, $A_2$, and $B_1$,
we can choose coefficients $m_{j,k}$ such that
\begin{eqnarray}
Z_{j}= m_{j,1}A_1+m_{j,2}A_2+m_{j,3}B_1 \Label{E30}.
\end{eqnarray}
Now, we assume that Eve attacks one of the channels 
$\{ e(5),\cdots, e(11)\}$, which is denoted by $e_A$. 
That is, Eve intercepts the channel $e_A$, keeps the information in $e_A$ on her hand,
and sends a new information $E_1$ though $e_A$. 
In this case, we denote the resultant values by $Z_{j}'$, which are
determined by $A_1$, $A_2$, $B_1$, and $E_1$.
Since the coefficients are different from $m_{j,k}$, 
we can choose other coefficients $m_{j,k}'$ such that
\begin{eqnarray}
Z_{j}'= m_{j,1}'A_1+m_{j,2}'A_2+m_{j,3}'B_1+m_{m,4}'E_1 \Label{O01}.
\end{eqnarray}
We should remark that the coefficient $\{ m_{i,j}' \}_{ij}$ depends on the choice of $e_A$, 
that is, the edge that Eve attacks. 
We can easily show the security of this classical network coding against this attack of Eve.

Here, we should remark the relation with existing works.
Indeed, the paper \cite{Agarwal} discussed classical secure network coding for multiple unicast scenario 
with butterfly network. However, they did not consider our code because they did not introduce shared randomness.

%For the latter discussion, we generalize the above conversion.
%When we set the initial values of $Z_5, \ldots, Z_{13}$ independently.
%the above conversions are generalized as
%\begin{eqnarray}
%\begin{split}
%Z_5 &\to 2A_1+B_1+Z_5, \quad 
%Z_6 \to 2A_2+B_1+Z_6 \\
%Z_7 &\to A_1+B_1+Z_7, \quad
%Z_8 \to A_2+B_1+Z_8 \\
%Z_9 &\to Z_5+Z_6+Z_9, \quad
%Z_{10} \to Z_9+Z_{10}\\
%Z_{11} &\to Z_9+Z_{11}, \quad
%Z_{12} \to \frac{1}{2}Z_{11}-Z_{8}+Z_{12} \\
%Z_{13} &\to \frac{1}{2}Z_{10}-Z_{7}+Z_{13}.
%\end{split}\Label{E32}
%\end{eqnarray}
%That is, the original conversion is the special case with the initial value $Z_5= \ldots= Z_{13}=0$.

\section{Quantum protocol}\Label{S3}
Based on the above classical protocol, we make a protocol to transmit quantum states
by the means of the idea used in \cite{Kobayashi2009,Kobayashi2010,Kobayashi2011};
our method is generalized from methods \cite{Kobayashi2009,Kobayashi2010,Kobayashi2011}
in order to treat classical shared randomness on the quantum network coding.

Our problem is formulated as transmission of quantum states in the Hilbert spaces 
${\cal H}_1$ and ${\cal H}_2$ 
to the output quantum systems 
${\cal H}_{12}$ and ${\cal H}_{13}$ that correspond to the edges $e(12)$ and $e(13)$, respectively,
where these quantum systems are spanned by the Z-basis $\{|a\rangle\}_{a=0}^{p-1}$.
Since edges $e(5), \ldots, e(11) $ correspond to quantum channels,
we assign them to the the same-dimensional quantum system ${\cal H}_5, \ldots, {\cal H}_{11}$, respectively.
Since the shared randomness $B_1$ and $B_2$ are classical information,
the edges $e(3), e(4), e(14), e(15)$ correspond to classical systems.
The variable $B_1$ takes values in $\FF_p$, but the variable $B_2$ takes in $\FF_p^2$.
So, while the edges $e(3)$ and $ e(4)$ send one element of $\FF_p$ in one time,
the edges $e(14), e(15)$ send $2$ elements of $\FF_p$ in one time.

Our protocol is given as Protocol \ref{protocol1}, which is composed of four steps. 
The unitaries used in Step 3 are given as follows.
\begin{eqnarray*}
U_5(b_1):=& \sum_{a_1,z_5}|a_1\rangle_1| 2a_1+b_1+z_5\rangle_5 ~_1\langle a_1 | ~_5\langle z_5| \\
U_6(b_1):=& \sum_{a_2,z_6}|a_2\rangle_2| 2a_2+b_1+z_6\rangle_6 ~_2\langle a_2 | ~_6\langle z_6| \\
U_7(b_1):=& \sum_{a_1,z_7}|a_1\rangle_1| a_1+b_1+z_7\rangle_7 ~_1\langle a_1 | ~_7\langle z_7| \\
U_8(b_1):=& \sum_{a_2,z_8}|a_2\rangle_2| a_2+b_1
+z_5\rangle_8 ~_2\langle a_2 | ~_8\langle z_8| ,
\\
%\end{eqnarray}
%and
%\begin{eqnarray}
U_9:=& \sum_{z_5,z_6,z_9}|z_5\rangle_5|z_6\rangle_6| z_5+z_6+z_9\rangle_9 
 ~_5\langle z_5| ~_6\langle z_6|~_9\langle z_9|\\
U_{10}:=& \sum_{z_9,z_{10}} |z_9\rangle_1| z_9+z_{10}\rangle_{10} ~_9\langle z_9 | ~_{10}\langle z_{10}| \\
U_{11}:=&\sum_{z_9,z_{11}} |z_9\rangle_1| z_9+z_{11}\rangle_{11} ~_9\langle z_9 | ~_{11}\langle z_{11}| \\
U_{12}:=& \sum_{z_8,z_{11},z_{12}}|z_8\rangle_8|z_{11}\rangle_{11}| \frac{1}{2}z_{11}- z_{8}+z_{12}\rangle_{12} \\
 &\quad~_8\langle z_8| ~_{11}\langle z_{11}|~_{12}\langle z_{12}|\\
U_{13}:=& \sum_{z_7,z_{10},z_{13}}|z_7\rangle_7|z_{10}\rangle_{10}|\frac{1}{2} z_{10}- z_{7}+z_{13}\rangle_{13} \\
 &\quad~_7\langle z_7| ~_{10}\langle z_{10}|~_{13}\langle z_{13}|.
 \end{eqnarray*}
 The $X$-basis used in Step 3 is given as
$|\phi_b\rangle:= \frac{1}{\sqrt{p}} \sum_{a=0}^{p-1} \omega^{ ab }|a\rangle$, where 
$\omega:=e^{2 \pi i/p}$.
The phase shift operator used in Step 4 is defined as
$\sX:=\sum_{a=0}^{p-1}\omega^a |a\rangle \langle a|$.

\begin{Protocol}                  
\caption{Secure network coding Protocol for butterfly network }         
\label{protocol1}      
\begin{algorithmic}
\STEPONE[Initialization]  
The system ${\cal H}_i$ is prepared on  the vertex $u$ with $e(i)=(u,v)$ for $i\ge 5$.
The systems ${\cal H}_1$ and ${\cal H}_2$ are prepared on the vertices $V_1$ and $V_2$, respectively.
We set the states on ${\cal H}_1$ and ${\cal H}_2$ to the states to be sent.
We set the states on ${\cal H}_{5}, \ldots, {\cal H}_{13}$ to be $|0\rangle$.

\STEPTWO[Transmission]  
The time counter starts from time $5$.
At time $i$ with $5 \le i \le 13$, we apply unitary $U_i$ on node $u$ with $e(i)=(u,v)$ based on the shared randomness $B_1=b_1$. 
Then, we send ${\cal H}_i$ to the node $v$ through the quantum channel $e(i)$. 

\STEPTHREE[Measurement on $X$-basis]  
We measure the systems 
${\cal H}_1, {\cal H}_2, {\cal H}_5, \ldots, {\cal H}_{13}$
with the $X$-basis $\{ |\phi_k\rangle\}_{k=0}^{p-1}$,
and obtain the outcomes $C_1,C_2,C_5 \ldots, C_{11}$. 
The outcomes $C_1, C_2,C_5\ldots, C_{11}$ 
are sent to the sink nodes $V_5$ and $V_6$ via public channels.
Only the outcomes $C_{10},  C_{11}$ 
are exchanged between the sink nodes $V_5$ and $V_6$
by using the shared randomness $B_2$.\

\STEPFOUR[Recovery]  
Based on the outcomes $C_1, C_2, C_5\ldots, C_{11}$,
The sink node $V_5$ applies the unitary 
$\sX^{-\sum_{k \in {\cal E}} C_k m_{k,2}}$ on ${\cal H}_{13}$.
The sink node $V_6$ applies 
the unitary 
$\sX^{-\sum_{k \in {\cal E}} C_k m_{k,1}}$ on ${\cal H}_{12}$,
where ${\cal E}:=\{1,2,5,6,\ldots,11 \}$.
\end{algorithmic}
\end{Protocol}

%\section{Check without Eve}

Now, we prove that the protocol properly transmits quantum states if there is no attack.
For this purpose,
%To verify whether the protocol properly transmits quantum states without any attack,
we input entanglement halves in ${\cal H}_1, {\cal H}_2$.
Let $\tilde{\cal H}_1, \tilde{\cal H}_2$ be the reference systems of ${\cal H}_1, {\cal H}_2$.
We prepare the maximally entangled state
$|\Phi\rangle_{\tilde{1},1}|\Phi\rangle_{\tilde{2},2}$, where
$|\Phi\rangle:=\frac{1}{\sqrt{p}}\sum_{a=0}^{p-1} |a,a\rangle$.
When the shared randomness $B_1$ is $b_1$,
the resultant state at after Step 2 is
\begin{eqnarray}
\frac{1}{p}\sum_{a_1,a_2} 
|a_1,a_2\rangle_{\tilde{1},\tilde{2}} %\nonumber \\
&| m_{1,1}a_1+m_{1,2}a_2+m_{1,3}b_1,\nonumber \\
&~ m_{2,1}a_1+m_{2,2}a_2+m_{2,3}b_1,\nonumber \\
& m_{5,1}a_1+m_{5,2}a_2+m_{5,3}b_1,\nonumber \\
&\ldots, \nonumber \\
&~m_{11,1}a_1+m_{11,2}a_2+m_{11,3}b_1,\nonumber \\
&~a_1,a_2\rangle_{5, \ldots, 13}. \label{O03}
\end{eqnarray}
%Notice that $m_{12,1}=1,m_{12,0}=0,m_{13,1}=0,m_{13,0}=1$.

When we obtain the outcomes $c_1, \ldots, c_{11}$ at Step 3,
the resultant state is
\begin{eqnarray}
&\frac{1}{p}\sum_{a_1,a_2} 
\omega^{\sum_{k \in {\cal E}} c_k (m_{k,1}a_1+m_{k,2}a_2+m_{k,3}b_1)  }
|a_1,a_2\rangle_{\tilde{1},\tilde{2}}
|a_1,a_2\rangle_{12, 13} \nonumber\\
=&
\sX_{12}^{\sum_{k \in {\cal E}} C_k m_{k,1}}
\sX_{13}^{\sum_{k \in {\cal E}} C_k m_{k,2}}
\omega^{\sum_{k \in {\cal E}} C_k m_{k,3}b_1  }
|\Phi\rangle_{\tilde{1},12}
|\Phi\rangle_{\tilde{2},13}.
\end{eqnarray}
Therefore, the resultant state after Step 4 is
$\omega^{\sum_{k \in {\cal E}} C_k m_{k,3}b_1  }
|\Phi\rangle_{\tilde{1},12}
|\Phi\rangle_{\tilde{2},13}$, which is the same as
$|\Phi\rangle_{\tilde{1},12}|\Phi\rangle_{\tilde{2},13}$ nevertheless the value of $b_1$.

\section{Security analysis}\Label{S4}
In this section, we prove the secrecy of the protocol against Eve's attack under the assumption that Eve does not know the shared randomness $B_1$ and $B_2$. 
We consider the situation that Eve attacks the quantum channel $e_A$,
where $e_A = e(j_E)$ with $5\le j_E \le 11$. 
In this situation, the most general attack of Eve can be described as follows:
In Step 2 of the protocol, 
Eve intercepts  $e_A$, and applies
a quantum operation $\Lambda _E$ defined from ${\cal H}_{j_E}$ 
to ${\cal H}_E \otimes {\cal H}_{j_E}$. 
Then, Eve keeps ${\cal H}_E$ and sends back ${\cal H}_{j_E}$ to 
the quantum channel $e_A$.
Further, in Step 3, Eve can access the information sent through the public channels 
except $C_{10}$ and $C_{11}$, which are encoded by means of $B_2$.

Here, we introduce notations: $M'$ is the matrix composed of $\{ m_{i,j}'\}_{i,j}$ 
with 1st, 2nd, 5th, $\cdots$,13th raws and $1-4$th columns.
Remember that the entries $m_{i,j}'$ are introduced in \eqref{O01}. 
Let $\lambda (a,b,x,y)$ be an operator on ${\cal H}_E$ defined by 
\begin{equation}
\lambda (a,b,x,y):=\left( I_E\otimes \langle x |_{j_E} \right) \cdot \Lambda_E \left (| a \rangle
\langle b | \right ) \cdot \left ( I_E \otimes | y \rangle _{j_E}  \right ) .
\end{equation}
Then, it satisfies 
$\Lambda _E\left ( | a\rangle \langle b| \right ) 
= \sum _{x,y} \lambda (a,b,x,y) \otimes | x \rangle \langle y |$, where 
$0 \le a,b,x,y \le p-1$. 
We further define an operator $\sigma _b$ on ${\cal H}_E$  depending on $0\le b \le p-1$ as 
\begin{eqnarray}
\sigma_b:= \sum  _{a=0}^{p-1}\lambda(a,a,b,b) =\left( I_E\otimes \langle b |_{j_E} \right) \cdot \Lambda_E \left ( I_{j_E} \right ) \cdot \left ( I_E \otimes | b \rangle _{j_E}  \right ).
\end{eqnarray}
From the definition of the operator $\lambda (a,b,x,y)$,
$\sigma_b$ is a positive operator and depends on the choice of $\Lambda_E$. % and $e_A$.

Our first observation is that we can judge the secrecy of the input state only from  the reduced density matrix after the protocol on Eve's system and the reference system,
which does not depend on whether we apply the recovery operation in the step 4 or not. Hence, we omit step 4 in our calculation for the security proof. 

First, the resultant state after Step 2, which was Eq.(\ref{O03}) in the presence of Eve, 
is proportional to 
\begin{eqnarray}
& \sum_{a_1, a_2,a_1',a_2',b_1,e_1,e_1'}   | a_1,a_2 \rangle \langle a_1',a_2' |_{\tilde{1},\tilde{2}} 
\nonumber \\ & \otimes  | M' (a_1,a_2,b_1,e_1)^T \rangle \langle M' (a_1',a_2',b_1,e_1')^T |_{1,2,5,\cdots,13}
\nonumber \\
& \otimes \lambda \left (\vec{m}_E \cdot (a_1,a_2,b_1)^T, \vec{m}_E \cdot (a_1',a_2',b_1)^T,
e_1,e_1' \right), \label{O10}
\end{eqnarray}
where $\vec{m}_E := (m_{j_E,1}, m_{j_E,2}, m_{j_E,3})$.

In Step 3, we separately consider the measurements on the systems ${\cal H}_{10} \otimes  {\cal H}_{11}$ and 
the measurements on the remaining systems. 
Since Eve does not know the outcomes $C_{10}$ and $C_{11}$,
from Eve's viewpoint, the measurements on the systems ${\cal H}_{10} \otimes  {\cal H}_{11}$
is equivalent to just tracing out ${\cal H}_{10} \otimes  {\cal H}_{11}$.
Further, since we omit the step 4, the systems ${\cal H}_{12} \otimes {\cal H}_{13}$ is not implemented any operation after the step 2. Hence, we can trace out  ${\cal H}_{12} \otimes {\cal H}_{13}$ as well.
So, we trace out ${\cal H}_{10} \otimes  {\cal H}_{11} \otimes {\cal H}_{12} \otimes {\cal H}_{13}$ 
from our calculation of the states. 
This calculation of the tracing-out depends on the choice of $e_A$. 
But the results of the calculation can be written in the same form and the state on the remaining systems before the measurement is proportional to 
\begin{eqnarray}
& \sum_{a_1, a_2,b_1,e_1}   | a_1,a_2 \rangle \langle a_1,a_2 |_{\tilde{1},\tilde{2}} 
\nonumber \\ & \otimes  | M'' (a_1,a_2,b_1,e_1)^T \rangle \langle M'' (a_1,a_2,b_1,e_1)^T |_{1,2,5,\cdots,9}
\nonumber \\
& \otimes \lambda \left (\vec{m}_E \cdot (a_1,a_2,b_1)^T, \vec{m}_E \cdot (a_1,a_2,b_1)^T,
e_1,e_1 \right),\label{O11}
\end{eqnarray}
where the matrix $M''$ is the submatrix of $M'$ derived by removing  $10-13$th raws. 
Here, we only present the derivation of the above equation for the case $e_A=e(7)$; 
we can similarly derive the equation in the other case. In the case $e_A=e(7)$, 
we derive 
\begin{eqnarray}
&(Z_{10},Z_{11},Z_{12},Z_{13}) \nonumber \\
=&(2a_1+2a_2+2b_1,2a_1+2a_2+2b_1, a_1, a_1+a_2+b_1-e_1).
\end{eqnarray}
Hence, the state derived by tracing out ${\cal H}_{10} \otimes  {\cal H}_{11} \otimes {\cal H}_{12} \otimes {\cal H}_{13}$ from Eq.(\ref{O10}) is proportional to 
\begin{eqnarray}
& \sum_{a_1, a_2,a_1',a_2',b_1,e_1,e_1'}   | a_1,a_2 \rangle \langle a_1',a_2' |_{\tilde{1},\tilde{2}} 
\nonumber \\ & \otimes  | M'' (a_1,a_2,b_1,e_1)^T \rangle_{1,\cdots,9} \langle M'' (a_1',a_2',b_1,e_1')^T |_{1,\cdots,9}
\nonumber \\
& \otimes \lambda \left (\vec{m}_E \cdot (a_1,a_2,b_1)^T, \vec{m}_E \cdot (a_1',a_2',b_1)^T,
e_1,e_1' \right) \nonumber \\
& \delta(2a_1+2a_2+2b_1,2a_1'+2a_2'+2b_1) \nonumber \\ &\delta(2a_1+2a_2+2b_1,2a_1'+2a_2'+2b_1) \nonumber \\ 
&\delta(a_1,a_1')\delta(a_1+a_2+b_1-e_1, a_1'+a_2'+b_1-e_1'). \label{O12}
\end{eqnarray}
It is easy to check the above equation is equal to Eq.(\ref{O11}).

Without losing generality, we assume that Eve writes all the measurement outcomes 
that  she derived in Step 3 into the Hilbert space ${\cal H}_\nu$ on Eve's hand.
Then, the state on $\tilde{\cal H} _1 \otimes \tilde{\cal H} _2 \otimes {\cal H}_E 
\otimes {\cal H}_\nu$ after the protocol  is proportional to 
\begin{eqnarray}
& \sum_{a_1, a_2,b_1,e_1, \vec{C}}   | a_1,a_2 \rangle \langle a_1,a_2 |_{\tilde{1},\tilde{2}} 
 \otimes | \vec{C} \rangle \langle \vec{C} |_{\nu} \nonumber \\
& \otimes \lambda \left (\vec{m}_E \cdot (a_1,a_2,b_1)^T, \vec{m}_E \cdot (a_1,a_2,b_1)^T,
e_1,e_1 \right),\label{O13}
\end{eqnarray}
where $\vec{C}=(C_1,C_2,C_5,\cdots,C_9)$ .
Since $m_{j_E,3} \neq 0$ for any choice of $e_A$, for fixed $a_1$ and $a_2$, $\lambda$ satisfies
\begin{eqnarray}
\sum _{b_1} \lambda \left (\vec{m}_E \cdot (a_1,a_2,b_1)^T, \vec{m}_E \cdot (a_1,a_2,b_1)^T,
e_1,e_1 \right)=\sigma_{e_1}. \label{O14}
\end{eqnarray}
From Eqs. (\ref{O13}) and (\ref{O14}), the state on  $\tilde{\cal H} _1 \otimes \tilde{\cal H} _2 \otimes {\cal H}_E 
\otimes {\cal H}_\nu$  after the protocol  is proportional to 
\begin{eqnarray}
\left( \sum_{e_1} \sigma_{e_1} \right ) \otimes I_{\tilde{1} \tilde{2}}\otimes I _\nu. 
\end{eqnarray} 
Hence, Eve's system after the protocol is independent from the reference system.
This completes the proof of the secrecy of our protocol. 
Therefore, our procol is secure from Eve's attack any one of the edges 
$e(5),\cdots, e(11)$.

At the last part of this section, we show the necessity of the shared randomness $B_2$.
As we have explained, the corresponding classical network coding on the butterfly network does not have a shared randomness corresponding to $B_2$. 
On the other hand, in our protocol, we use $B_2$ taking  value in $\FF _p^2$, 
which is equal to two elements of $\FF_p$ and used to send $C_{10}$ and $C_{11}$, securely. 
Here, we consider the situation that only $C_{11}$ is encoded by shared randomness,
and Eve derives the information of $C_{10}$. This is the case when the size of $B_2$  
is $\FF_p$, which is smaller than that of the present protocol.
It is possible to show that the protocol is not secure in this case, and there exists an attack  of Eve by which she can derive the information of the quantum states. 
Suppose Eve attacks the channel $e(11)$ by $\Lambda_E$ that is defined as 
$\Lambda _E(|a \rangle \langle b| )  := 
|a \rangle \langle b| _E \otimes |\phi _0 \rangle \langle \phi_0|_{11}$; 
that is, Eve just keeps the state on $e(11)$ on her hands, 
and  sends $|\phi_0\rangle$ back to the channel $e(11)$.
By the straightforward calculation, we can show that the state on
$\tilde{{\cal H}}_1\otimes \tilde{{\cal H}}_1\otimes {\cal H}_E$ after the protocol is 
 \begin{eqnarray}
 \frac{1}{p^3} \sum_{a_1,a_2,a_1',b_1} & |a_1,a_2\rangle \langle a_1' a_2|_{\tilde{1}\tilde{2}} 
 \nonumber \otimes |2a_1+2a_2+2b_1 \rangle \langle 2a_1'+2a_2+2b_1|_E.
\end{eqnarray}
Hence, Eve's system is not independent from the reference system,
and Eve can derive the information of the quantum state to be sent in this protocol. 
Therefore, for the security of the protocol, we need to hide both $C_{10}$ and $C_{11}$ from Eve, and we need to use the extra shared randomness $B_2$ taking the value in $\FF_p^2$.

\section{Conclusion}\Label{S5}
We have proposed secure quantum network coding on the butterfly network in the multiple unicast setting based on a secure classical network coding.
This protocol certainly transmits quantum states when there is no attack.
We also have shown the secrecy even when 
the eavesdropper wiretaps one of the channels in the butterfly network,
which does not require any additional verification protocol.

Our security proof can be extended to a more general situation \cite{Gene}.
That is, when the corresponding classical network code satisfies the robustness \cite{Ho2008,Jaggi2008,Nutman2008,Yu2008}
in addition to the secrecy, 
we can prove the security similar to this paper.
Here, we need to discuss the secrecy even when the eavesdropper contaminates a part of information as well as eavesdrops the part of information like \cite{HOKC}
while the conventional security papers 
\cite{Cai2002,Cai06,Bhattad2005,Liu2007,Rouayheb2007,Harada2008,SK,Cai2011,Cai2011a,Matsumoto2011,Matsumoto2011a,KMU} 
for network coding discussed only the secrecy only when 
the eavesdropper eavesdrops a part of information but does not contaminate the part of information.
Since this kind of general analysis requires much more pages, we cannot discuss it in this paper.
So, our next paper discusses this kind of security analysis for quantum network coding.

%We will discuss this kind of generalization in a future paper \cite{OKH}.

\section*{Acknowledgments}
The authors are very grateful to 
Professor Ning Cai and Professor Vincent Y. F. Tan
for helpful discussions and comments.
The works reported here were supported in part by 
the JSPS Grant-in-Aid for Scientific Research 
(A) No. 23246071,
(C) No. 16K00014,  (C) No.  26400400, and (B) No. 16KT0017,
the Okawa Research Grant,
and Kayamori Foundation of Informational Science Advancement.

%\bibliographystyle{ieeetr}
%\bibliography{secure_network_coding}

\section*{References}

\begin{thebibliography}{99}

%\bibitem{BCW99}
% H. R. Buhrman, R. Cleve, and A. Wigderson, ``{Quantum vs. Classical Communication and Computation},''
%in Proceedings of the 30th Annual ACM Symposium on Theory of Computing (ACM, New York, NY, USA, 1999), pp. 63-68. 1999
 
%\bibitem{R99} R. Raz, ``{Exponential Separation of Quantum and Classical Communication Complexity},'' in Proceedings of the Thirty-first Annual ACM Symposium on Theory of Computing, STOC 1999, pp. 358--367. ACM, New York, NY, USA, 1999. 

\bibitem{Cai2002}
N. Cai and R.~Yeung, 
``{Secure network coding},'' 
in {\em Proceedings of 2002 IEEE International Symposium on Information Theory (ISIT)},
pp. 323, 2002.

\bibitem{Cai06}
N. Cai and R. W. Yeung, 
``Network error correction, Part 2: Lower bounds,'' 
{\em Commun. Inf. and Syst.}, vol. 6, no. 1, 37--54, Jan. 2006.

\bibitem{Bhattad2005}
K.~Bhattad, S.~Member, and K.~R. Narayanan, 
``{Weakly Secure Network Coding},''
in {\em First Workshop on Network Coding, Theory, and Applications}, (Riva del Garda), 2005.

\bibitem{Liu2007}
R.~L.~R. Liu, Y.~L.~Y. Liang, H.~Poor, and P.~Spasojevic, 
``{Secure Nested Codes for Type II Wiretap Channels},'' 
{\em 2007 IEEE Information Theory Workshop}, pp.~337--342, 2007.

\bibitem{Rouayheb2007}
S.~Y.~E. Rouayheb and E.~Soljanin, 
``{On Wiretap Networks II},'' 
in {\em Proceedings of 2007 IEEE International Symposium on Information Theory (ISIT)},
pp.~551--555, 2007.

\bibitem{Harada2008}
K.~Harada and H.~Yamamoto, ``{Strongly Secure Linear Network Coding},'' 
{\em  IEICE transactions on Fundamentals of Electronics, Communications and Computer Sciences}, 
vol.~E91-A, no.~10,~2720--2728, 2008.

\bibitem{SK}
D. Silva and F. R. Kschischang, 
``Security for wiretap networks via rank metric codes,'' 
in {\em Proceedings of 2008 IEEE International Symposium on Information Theory (ISIT)},
pp. 176 -- 180, 2008.

\bibitem{Cai2011}
N.~Cai and T.~Chan, ``{Theory of Secure Network Coding},'' 
{\em Proceedings of  the IEEE}, vol.~99, ~421--437, 2011.

\bibitem{Cai2011a}
N.~Cai and R.~W. Yeung, 
``{Secure Network Coding on a Wiretap Network},''
{\em IEEE Transactions on Information Theory},  vol. 57, no. 1, 424 -- 435, 2011.

\bibitem{Matsumoto2011}
R.~Matsumoto and M.~Hayashi, 
``{Secure Multiplex Network Coding},'' 
{\em 2011 International Symposium on Networking Coding} (2011):
DOI: 10.1109/ISNETCOD.2011.5979076.

\bibitem{Matsumoto2011a}
R.~Matsumoto and M.~Hayashi, 
``{Universal Secure Multiplex Network Coding with Dependent and Non-Uniform Messages},'' 
Accepted for publication in \emph{IEEE Trans.\ Inform.\ Theory};
{\em Arxiv preprint}, arXiv: 1111.4174 (2011).

\bibitem{KMU}
J. Kurihara, R. Matsumoto, and T. Uyematsu,
``Relative generalized rank weight of linear codes and its applications to network coding,''
{\em IEEE Transactions on Information Theory}, 
vol.~61, no.~7, 3912--3936, 2013.

\bibitem{Hayashi2007}
M.~Hayashi, K.~Iwama, H.~Nishimura, R.~Raymond, and S.~Yamashita, 
``{Quantum Network Coding},'' 
in {\em STACS 2007 SE - 52} (W.~Thomas and P.~Weil, eds.), vol.~4393 of 
{\em Lecture Notes in Computer Science}, pp.~610--621, Springer Berlin Heidelberg, 2007.

\bibitem{PhysRevA.76.040301}
M.~Hayashi, 
``{Prior entanglement between senders enables perfect quantum network coding with modification},'' 
{\em Phys. Rev. A}, vol.~76, no.~4,~40301, 2007.

\bibitem{Kobayashi2009}
H.~Kobayashi, F.~{Le Gall}, H.~Nishimura, and M.~R\"{o}tteler, ``{General
  Scheme for Perfect Quantum Network Coding with Free Classical
  Communication},'' in {\em Automata, Languages and Programming SE - 52}
  (S.~Albers, A.~Marchetti-Spaccamela, Y.~Matias, S.~Nikoletseas, and
  W.~Thomas, eds.), 
  vol.~5555 of 
  {\em Lecture Notes in Computer Science},
  pp.~622--633, Springer Berlin Heidelberg, 2009.

\bibitem{Leung2010}
D.~Leung, J.~Oppenheim, and A.~Winter, 
``{Quantum Network Communication; The Butterfly and Beyond},'' 
{\em IEEE Transactions on Information Theory}, vol.~56, no.~7,~3478--3490, 2010.

\bibitem{Kobayashi2010}
H.~Kobayashi, F.~{Le Gall}, H.~Nishimura, and M.~Rotteler, 
``{Perfect quantum network communication protocol based on classical network coding},'' 
in {\em Proceedings of 2010 IEEE International Symposium on Information Theory (ISIT)},
pp.~2686--2690, 2010.

\bibitem{Kobayashi2011}
H.~Kobayashi, F.~{Le Gall}, H.~Nishimura, and M.~Rotteler, ``{Constructing
  quantum network coding schemes from classical nonlinear protocols},'' 
in {\em Proceedings of 2011 IEEE International Symposium on Information Theory (ISIT)},
pp.~109--113, 2011.

\bibitem{PhysRevLett.101.060401}
G.~Chiribella, G.~M. D'Ariano, and P.~Perinotti, 
``Quantum circuit architecture,'' 
{\em Phys. Rev. Lett.}, vol.~101,~060401, 2008.

\bibitem{PhysRevA.80.022339}
G.~Chiribella, G.~M. D'Ariano, and P.~Perinotti, 
``Theoretical framework for quantum networks,'' 
  {\em Phys. Rev. A}, vol.~80,~022339, 2009.

\bibitem{Gottesman1999}
D.~Gottesman and I.~L. Chuang, 
``{Demonstrating the viability of universal quantum computation using teleportation and single-qubit operations},'' 
  {\em Nature}, vol.~402, pp.~390--393, 1999.

\bibitem{PhysRevLett.70.1895}
C.~H. Bennett, G.~Brassard, C.~Cr\'epeau, R.~Jozsa, A.~Peres, and W.~K. Wootters, 
  ``Teleporting an unknown quantum state via dual classical and einstein-podolsky-rosen channels,'' 
  {\em Phys. Rev. Lett.}, vol.~70,~1895--1899, 1993.

%\bibitem{CF} F. Cheng and  V. Y. F. Tan,
%``A Numerical Study on the Wiretap Network With a Simple Network Topology,''
%IEEE Transactions on Information Theory, Volume: 62, Issue: 5, 2481 - 2492, (2016) 

 \bibitem{ACLY}
R. Ahlswede, N. Cai, S. -Y. R. Li, and  R. W. Yeung,
``Network information flow,''
{\em IEEE Transactions on Information Theory},
vol. 46, no. 4, 1204 -- 1216, 2000.

\bibitem{Ho2008}
T.~H.~T. Ho, B.~L.~B. Leong, R.~Koetter, M.~Medard, M.~Effros, and D.~Karger,
``{Byzantine Modification Detection in Multicast Networks With Random Network Coding},'' 
{\em IEEE Transactions on Information Theory}, vol.~54, no.~6,~2798 -- 2803, 2008.

\bibitem{Jaggi2008}
S.~Jaggi, M.~Langberg, S.~Katti, T.~Ho, D.~Katabi, M.~Medard, and M.~Effros,
  ``{Resilient Network Coding in the Presence of Byzantine Adversaries},'' 
{\em IEEE Transactions on Information Theory}, vol.~54, no.~6, 2596--2603, 2008.

\bibitem{Nutman2008}
L.~Nutman and M.~Langberg, 
``{Adversarial models and resilient schemes for network coding},'' 
in {\em Proceedings of 2008 IEEE International Symposium on Information Theory (ISIT)},
pp.~171--175, 2008.

\bibitem{Yu2008}
Z.~Y.~Z. Yu, Y.~W.~Y. Wei, B.~Ramkumar, and Y.~G.~Y. Guan, 
``{An Efficient Signature-Based Scheme for Securing Network Coding Against Pollution Attacks},'' 
  {\em IEEE INFOCOM 2008 - The 27th Conference on Computer Communications}, 2008.

\bibitem{Agarwal}
G. K. Agarwal, M. Cardone, and C. Fragouli,
``On (Secure) Information flow for Multiple-Unicast Sessions: Analysis with Butterfly Network,''
arXiv:1606.07561 (2016).



\bibitem{HOKC}
M. Hayashi, M. Owari, G. Kato, and N. Cai, 
%``Secrecy and Robustness for Active Attack in Secure Network Coding and its Application to Network Quantum Key Distribution,''
arXiv: 1703.00723 (2017); 
Acceptd for 2017 IEEE International Symposium on Information Theory (ISIT),  Aachen (Germany), 25-30 June 2017.

\bibitem{Gene}
M. Owari, G. Kato, and M. Hayashi, 
``Secure Quantum Network Coding for General Multiple Unicast Network,''
In preparation.


%\bibitem{OKH} M. Owari, G. Kato, and M. Hayashi, in preparation.

\end{thebibliography}

\end{document}